\documentclass[10pt,aps,pra,showkeys,twocolumn,superscriptaddress]{revtex4-2}
\usepackage{amsmath, amssymb}
\usepackage{mathrsfs}
\usepackage{array}
\usepackage{booktabs}
\usepackage{tabu}
\usepackage{dcolumn}
\usepackage{amsfonts}
\usepackage{float}
\usepackage{graphicx,color}
\usepackage[colorlinks={true}]{hyperref}
\hypersetup{colorlinks=true,linkcolor=red,citecolor=blue,urlcolor=blue}
\usepackage{bm}
\usepackage{subcaption}
\usepackage{pstricks}
\usepackage{braket}
\graphicspath{{images/}}
\usepackage{caption}
\expandafter\let\csname equation*\endcsname\relax
\expandafter\let\csname endequation*\endcsname\relax
\usepackage{mathtools}
\usepackage[utf8]{inputenc}
\usepackage[T1]{fontenc}
\usepackage{mathptmx}
\usepackage[markup=underlined]{changes}
\usepackage{cleveref}
\usepackage{etoolbox}
\usepackage{multirow}
\usepackage{orcidlink}
\usepackage{listings}
\usepackage{xcolor}

\begin{document}

\title{Entanglement certification in bulk nonlinear crystals for degenerate and non-degenerate SPDC: spectral filter effects on transverse spatial correlations}

\author{Hashir~Kuniyil\orcidlink{0000-0003-0338-1278}} 
\email{hkuniyil@hbku.edu.qa}
\affiliation{Qatar Center for Quantum Computing, College of Science and Engineering, Hamad Bin Khalifa University, Doha, Qatar}
\author{Asad Ali\orcidlink{0000-0001-9243-417X}} 
\affiliation{Qatar Center for Quantum Computing, College of Science and Engineering, Hamad Bin Khalifa University, Doha, Qatar}
\author{Saif Al-Kuwari\orcidlink{0000-0002-4402-7710}}

\date{\today}

\begin{abstract}
Spatial correlations of photon pairs from spontaneous parametric down-conversion (SPDC) underpin quantum imaging and entanglement certification. We present the first systematic study of spectral filter bandwidth effects on transverse spatial correlations in bulk Type-I BBO for degenerate and non-degenerate configurations. In the far field, the degenerate conditional momentum width is pump-limited and filter-invariant, while non-degenerate configurations exhibit monotonic growth in both marginal and conditional momentum widths — with the walk-off axis $\approx 100$ times more sensitive than the non-walk-off axis. In the near field, we identify a previously unreported flat-dip-rise profile: the conditional position width narrows by $\approx 10\%$ at an optimal bandwidth $\Delta_\mathrm{dip} \approx 1.35\,\Delta\lambda_\mathrm{SPDC}$ before rising due to geometric displacement. When the filter is placed on the idler arm, the dip shifts by the exact factor $(\lambda_i/\lambda_s)^2$. Both results are universal for any non-degenerate SPDC source, requiring only a finite crystal length, $d\theta/d\lambda \neq 0$, and incoherent spectral averaging. The Reid EPR uncertainty product is consistently smaller on the walk-off axis — a structural advantage of bulk birefringent geometry absent in quasi-phase-matched sources. The optimal filter bandwidth $\Delta_F = \Delta_\mathrm{dip}$ is determined entirely by the intrinsic phase-matching bandwidth of the crystal and is directly readable from the X-entanglement spectral width of the source.
\end{abstract}

\maketitle
\section{Introduction}

Imaging schemes based on correlated and entangled photon pairs have become central to quantum sensing and imaging \cite{howell2004realization, pirandola2018advances, brida2010experimental, morris2015imaging, magana2013compressive, brambilla2008high, defienne2019quantum, lemos2014quantum, lahiri2015theory, berchera2019quantum}, offering advantages such as enhanced noise resilience \cite{kuniyil2020object, aspden2015photon, morris2015imaging, kuniyil2022noise, gregory2021noise}, flexible wavelength operation \cite{paterova2020quantum, aspden2015photon, moreau2018ghost}, and
access to nonclassical correlation observables
\cite{toninelli2019resolution, zhang2024quantum, he2023quantum, defienne2022pixel}. These schemes generally rely on photon pairs generated via spontaneous parametric down-conversion (SPDC) in a
nonlinear crystal \cite{brambilla2004simultaneous, klyshko1967coherent, burnham1970observation, hong1986experimental, paulk, boyd2020nonlinear,
bouwmeester1997experimental, kuniyil2021efficient, karan2020phase}, where strong correlations manifest in both transverse momentum and position spaces \cite{walborn2010spatial}.

In correlation-based imaging protocols such as quantum ghost imaging \cite{moreau2018ghost, moreau2018resolution, d2005resolution,
padgett2017introduction, aspden2013epr}, entanglement-assisted microscopy \cite{aspden2015photon, he2023quantum, hell1994breaking}, and imaging with undetected photons \cite{lemos2014quantum, fuenzalida2022resolution, lahiri2015theory}, the achievable spatial resolution is fundamentally limited by the strength of the transverse correlations between the photon pair \cite{d2005resolution}. In practical SPDC sources, the finite phase matching bandwidth — set by
the crystal length, pump-beam properties, and birefringence — introduces uncertainty in the one-to-one mapping between photon pairs, which in turn restricts the achievable resolution \cite{edgar2012imaging, bhattacharjee2022propagation, chan2007transverse, defienne2024advances}. Characterising how spatial correlations respond
to experimentally controllable input parameters is therefore of fundamental importance for biphoton-based quantum imaging.

The influence of crystal length and pump beam waist on SPDC spatial correlations is well established \cite{pires2011type, ramirez2013effects}. Schneeloch and Howell \cite{Schneeloch2016TransverseSPDC} showed through the biphoton birth-zone framework that the conditional position width scales as $\sqrt{L}$ — a
consequence of the phase matching sinc function — while the conditional momentum width is set exclusively by the pump beam waist, $\Delta q_{s|i} \approx 1/w_0$, independently of crystal length. These scalings were confirmed experimentally by Edgar et al.\ \cite{edgar2012imaging} and Howell et al.\ \cite{howell2004realization}, and have become the standard design relations for camera-based entanglement characterisation \cite{walborn2010spatial}.

Beyond these isotropic relations, additional structure arises in crystals where birefringent walk-off breaks the transverse symmetry. In Type-I nonlinear crystals such as BBO, the pump propagates as an extraordinary ray and the signal and idler as ordinary rays, producing a walk-off angle $\rho$ between the pump Poynting vector and the crystal axis. Fedorov et al.\ \cite{fedorov2007anisotropy} showed that this walk-off produces anomalously strong narrowing of the coincidence distribution on the walk-off axis in momentum space 

In SPDC wavelength selection, distinct design considerations arise depending on whether the source is operated at degeneracy or in a non-degenerate configuration. Degenerate SPDC has been extensively
studied, whereas the non-degenerate regime remains comparatively underexplored. 
As non-degenerate SPDC sources are increasingly important for wavelength-flexible quantum imaging, infrared--visible ghost imaging \cite{moreau2018ghost, moreau2018resolution}, and multimodal detection architectures, establishing design rules for this regime is essential.

The fundamental distinction between non-degenerate and degenerate SPDC in the context of spatial correlations is the intrinsic coupling between frequency and emission angle imposed by the phase matching condition: in non-degenerate operation, different signal wavelengths are phase-matched to different emission angles, therefore $d\theta/d\lambda \neq 0$. Gatti et al.\ \cite{gatti2009xentanglement} showed that this spatio-spectral coupling gives rise to non-factorable joint spatiotemporal correlations - called X-entanglement — whose structure differs fundamentally from the near-factorable degenerate case where $d\theta/d\lambda \approx 0$. Despite this foundational result, its quantitative consequences for experimentally accessible observables —
specifically, how spectral filter bandwidth independently shapes the conditional momentum and conditional position widths in the near- and far-field — have not been explicitly characterised in bulk birefringent crystals.

This paper addresses these gaps through the following contributions. \textit{First}, we characterise the effect of spectral filter bandwidth on the far-field (momentum-space) conditional and marginal widths in bulk Type-I BBO for degenerate and non-degenerate configurations. In the degenerate case, both widths are insensitive to filter bandwidth due to $d\theta/d\lambda \approx 0$. In non-degenerate SPDC, however, the marginal width broadens monotonically with filter bandwidth, following a power-law crossover model with sub-quadratic exponent $\beta \approx 1.2$--$1.35$, reflecting the sinc-shaped (rather than Gaussian) single-slice marginal distribution. We show explicitly that the conditional momentum width $\Delta q_{s|i}$ also grows with filter bandwidth in non-degenerate SPDC, in sharp contrast to the degenerate case where it remains pump-limited at $1/w_0$. The growth follows an exponential saturation model, and is strongly axis-dependent: the walk-off axis exhibits approximately 100 times greater variation in $\Delta q_{s|i}$ than the non-walk-off axis, driven by the spectral dependence of the walk-off spatial filter on the sum-momentum distribution \cite{dacosta2024epr}.

\textit{Second}, we characterise the near-field (position-space) conditional width and identify a previously unreported flat-dip-rise profile as a function of spectral filter bandwidth for all non-degenerate configurations. At narrow bandwidths the conditional width is filter-invariant and equal on both axes, following standard expression $\Delta x_{s|i}^{(0)} \approx \sqrt{0.488L\lambda_p/2\pi}$. As the filter widens beyond the intrinsic SPDC phase matching bandwidth $\Delta\lambda_\text{SPDC}$, the conditional width \emph{decreases} due to a Fourier-width mechanism: shorter-wavelength components within the filter passband carry a larger wavenumber $k_s$, producing a broader phase matching sinc in momentum space and a correspondingly narrower near-field distribution. The minimum occurs at $\Delta_\text{dip} \approx 1.35\,\Delta\lambda_\text{SPDC}$, yielding
an approximately 10\% improvement in conditional position width relative to the base value $\Delta x_{s|i}^{(0)}$. Beyond this optimum, the conditional width rises due to geometric displacement of near-field peaks from spectrally offset slices. When the filter is placed on the idler arm, the dip shifts by the exact factor $(\lambda_i/\lambda_s)^2$ — a parameter-free consequence of energy conservation, confirmed numerically to within 0.2\% and providing a direct experimental signature of spatio-spectral coupling.

\textit{Third}, combining both these effects, we evaluate the Reid EPR uncertainty product as the entanglement certification metric \cite{reid1989demonstration} and show that the near-field dip produces a direct improvement in certification strength at $\Delta_F = \Delta_\text{dip}$, while the degenerate case remains unaffected by filtering throughout. The walk-off axis consistently yields a lower uncertainty product than the non-walk-off axis in the large-pump regime, confirming the structural advantage of bulk birefringent geometry identified by Fedorov et al.\
\cite{fedorov2007anisotropy}. Together, these results provide the first analytical filter selection guidelines for non-degenerate quantum imaging in bulk birefringent crystals: the optimal filter bandwidth is $\Delta_F = \Delta_\text{dip}$, given directly by the intrinsic SPDC phase matching bandwidth of the crystal.

\section{Theory}
\subsection{SPDC in BBO crystals}
\label{Sec:Phase_matching}
In the SPDC process, higher frequency pump field (p) is down converted into lower frequency signals, conventionally called the signal (s) and idler (i) photons. This process is governed by energy and momentum conservation that require \cite{walborn2010spatial} $\omega_p = \omega_s + \omega_i$ and $k_p \approx k_s + k_i$, where $\omega$ is angular frequency and $k$ is the wavenumber. 

For a Type-I SPDC in a uniaxial crystal such as $\beta$-barium borate (BBO), the pump is extraordinarily polarized while the signal and idler are ordinarily polarized. Under the paraxial approximation (refer appendix.~(\ref{app_A}) for detailed derivation), the longitudinal phase-mismatch can be written as
\begin{equation}
    \Delta k_z \approx k_{p,z} - k_{s,z} - k_{i,z} - \frac{(q_s + q_i)^2}{2k_p} - (q_s+q_i) \tan \rho,
\end{equation}
where $q_s$ and $q_i$ are the transverse wavenumbers and $\rho$ is the pump walk-off angle found using Eq.~\ref{eq:walk-off}. The corresponding phase matching (PM) efficiency is given by
\begin{equation}
    \Psi(\lambda_s,\theta)\propto\operatorname{sinc}^2\left(\frac{\Delta k_z (\theta) L}{2}\right),
    \label{eq:sinc}
\end{equation}
where $\theta$ is the PM angle, and $L$ is the crystal length. The finite crystal length, therefore, defines a momentum bandwidth over which SPDC emission occurs, while birefringent walk-off introduces an asymmetry along the walk-off axis (we consider y-direction as walk-off axis throughout our analysis). 

In addition to nonlinear crystal properties, the transverse structure of the pump beam plays a central role in determining SPDC spatial correlations. For a Gaussian pump with waist $w_0$, the pump envelope, combined with the nonlinear optical properties (Eq. (\ref{eq:sinc})), determines the biphoton amplitude (see appendix.~\ref{app_B} for more details)
\begin{align}
    \psi(q_x,q_y) \propto \exp{\left(\frac{(q_s+q_i)^2w_0^2}{4}\right)} \operatorname{sinc}^2\left(\frac{\Delta k_z L}{2}\right).
    \label{eq3}
\end{align}
This condition enforces approximate momentum anti-correlation between the signal and idler photons and acts in conjunction with the biphoton amplitude to determine the overall joint angular distribution. This explains the fundamental understanding that larger pump waists lead to tighter transverse correlations (in momentum space), while tightly focused pumps broaden the conditional distributions \cite{howell2004realization, monken1998transfer}, such that the conditional momentum width follows $\Delta q\approx1/w_0$, where $w_0$ is the pump beam waist radius.
To study the effects of introducing a spectral filter, we model the biphoton state generated in the BBO crystal using the standard SPDC phase matching formalism with a Gaussian bandpass filter.
\begin{equation}
\small
I(q_{s,x}, q_{i,x}) \propto \int d\lambda_s \left| \Psi(q_{s,x}, q_{i,x}; \lambda_s) \right|^2 F_s(\lambda_s),
\label{eq4}
\end{equation}
where $F_s(\lambda_s) = \exp[-\frac{(\lambda_s-\lambda_{s0})^2}{2\sigma_\lambda^2}]$ is a Gaussian bandpass filter in the signal wavelength, with $\sigma_\lambda = \frac{\mathrm{FWHM}}{2\sqrt{2\ln 2}}$. Using our spectral-filter-based analysis, we tune the full width at half maximum (FWHM) bandwidth \(\Delta\lambda\) to examine its effect on conditional correlation and marginal distributions. In practice, this filter transmission function, similar to experiments, restricts the range of frequencies permitted for the signal photon (and correspondingly the idler wavelength via \(\lambda_i = \lambda_p - \lambda_s\)).
\begin{figure*}[htbp]
  \centering
  \includegraphics[width=0.9\textwidth]{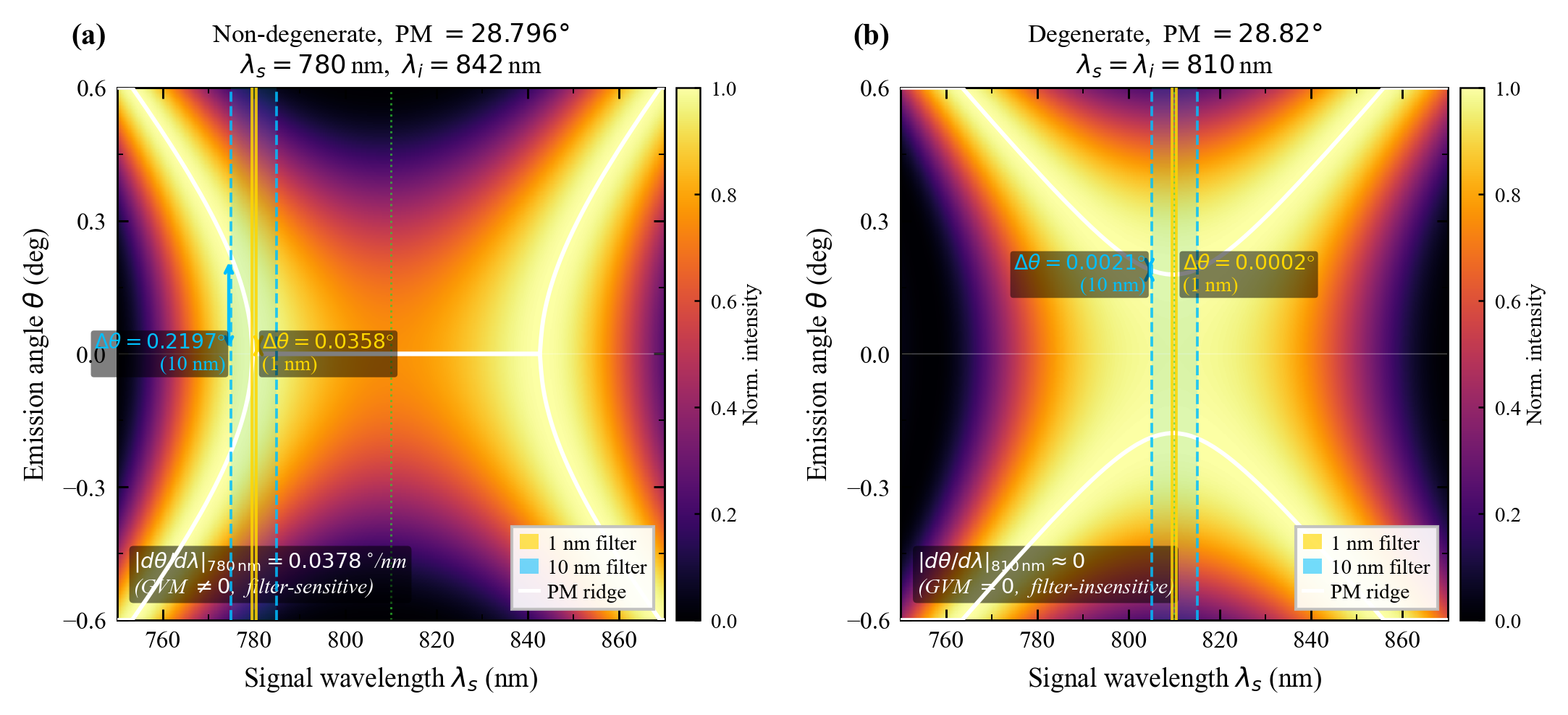}
  \caption{X-entanglement profile for (a) the non-degenerate case with $\lambda_s = 780\,\mathrm{nm}$ and $\lambda_i = 842\,\mathrm{nm}$, and (b) the degenerate case with $\lambda_s = \lambda_i = 810\,\mathrm{nm}$. Vertical lines indicate spectral bandwidths of 1 nm and 10 nm and the corresponding $\Delta\theta$ values for both cases. }
  \label{fig:fig1}
\end{figure*}
Eq.~(\ref{eq4}) yields the momentum JPD $I(q_{s,x},q_{i,x})$, which we use to understand the effect of the spectral filter on the conditional uncertainty in the momentum space. 

To obtain the position-space correlation $I(x_s,x_i)$, we perform a fast Fourier transform (FFT) of the filtered biphoton amplitude  which transfers the intensity distribution from momentum space, $(q_{s,x},q_{i,x})$, to the position space, $(x_s,x_i)$, following, 

\begin{equation}
\begin{aligned}
I(x_s, x_i)
  &= \int d\lambda_s \, F_s(\lambda_s)\\&
     \times\left| \iint dq_{s,x} \, dq_{i,x} \,
       \Psi(q_{s,x}, q_{i,x};\, \lambda_s)
       e^{i(q_{s,x} x_s + q_{i,x} x_i)} \right|^2 \\
\label{eq:position_Intensity}
\end{aligned}
\end{equation}

which is essentially the biphoton analog of an optical Fourier transform. (In experiment, this transformation can be done by imaging the crystal face onto a camera with a 1:$M$ magnification). To determine the position-space uncertainty ($\Delta x_{s|i}$) and momentum-space uncertainty ($\Delta q_{s|i}$), we used the method described in the following section-- Sec.~\ref{sec:estimatingUncertainity_parameter}.

\subsection{Spatial Uncertainty Parameters}
\label{sec:estimatingUncertainity_parameter}
To analyze the conditional correlations in quantum imaging, the standard way is to select a spatial point for the signal (or idler) and examine the distribution of its counterpart, the idler (or signal) \cite{defienne2024advances}. In our simulation, to estimate the spatial correlation, we build the full JPD function for the SPDC field in both the momentum space $I(q_s, q_i)$ (Eq. \ref{eq4}) and the position space $I(x_s, x_i)$ (Eq.~\ref{eq:position_Intensity}), and then calculated the inferred variance using the covariance formula.
\begin{equation} 
Var(q_i | q_s) = V_{q_i} - \frac{C_{q_i, q_s}^2}{V_{q_s}},
\label{eq:Cond_var}
\end{equation}
which is the standard Reid EPR estimator where $V$ stands for individual field variance, and $C$ is the covariance of the signal and idler fields. When this is applied in the x-direction (no walk-off axis), the JPD naturally develops a ridge near $q_{i,x} \approx -q_{s,x}$ due to the momentum anti-correlation. In the y-direction (walk-off axis), the JPD follows the stationary line 
$q_{i,y} \approx k_{y,p} - q_{s,y}$ (a shifted anti-correlation), this shift results from the  
($\Delta k_y, \text{term in } \Delta k_z$, Ref. Eq. (\ref{eq:DeltaKz})). We consider the JPD measured either in the far-field (momentum space) \(I_q(a_s,a_i),\ a\in\{q_x,q_y\}
\) or in the near-field (position space) \(I_x(a_s,a_i),\ a\in\{x,y\}\). On a rectangular grid defined by the points \(a_s[k]\) (\(k=1,\dots,N_s\)) and \(a_i[\ell]\) (\(\ell=1,\dots,N_i\)). We convert the intensity into a discrete probability table by normalizing with the grid measures following:
\begin{widetext}
\begin{equation}
P_{k\ell}
=\frac{
I\!\left(a_s[k],a_i[\ell]\right)\,\Delta a_s\,\Delta a_i
}{
\displaystyle
\sum_{k=1}^{N_s}\sum_{\ell=1}^{N_i}
I\!\left(a_s[k],a_i[\ell]\right)\,\Delta a_s\,\Delta a_i
},
\quad
\sum_{k,\ell} P_{k\ell}\,\Delta a_s\,\Delta a_i = 1.
\label{eq:prob_table}
\end{equation}
\end{widetext}
This implementation ensures that all subsequent moments are computed from a proper probability distribution and that the units remain consistent. For uniform grids, we can equivalently normalize by the plain sum, since constant \(\Delta a\) factors cancel.

\paragraph{First moments.}
The signal and idler means are
\begin{align}
\mu_s &\equiv \mathbb{E}[a_s]
= \sum_{k,\ell} a_s[k]\;P_{k\ell}\,\Delta a_s\,\Delta a_i, \notag\\
\mu_i &\equiv \mathbb{E}[a_i]
= \sum_{k,\ell} a_i[\ell]\;P_{k\ell}\,\Delta a_s\,\Delta a_i.
\label{eq:means}
\end{align}
\paragraph{Variances and the covariance.}
The marginal variances and the signal-idler covariance are
\begin{align}
V_s &\equiv \mathrm{Var}(a_s)
= \sum_{k,\ell} \bigl(a_s[k]-\mu_s\bigr)^2\,P_{k\ell}\,\Delta a_s\,\Delta a_i, \label{eq:var_s}\\
V_i &\equiv \mathrm{Var}(a_i)
= \sum_{k,\ell} \bigl(a_i[\ell]-\mu_i\bigr)^2\,P_{k\ell}\,\Delta a_s\,\Delta a_i, \label{eq:var_i}
\end{align}
\begin{align}
C_{si} &\equiv \mathrm{Cov}(a_s,a_i) \notag\\
&= \sum_{k,\ell} 
\bigl(a_s[k]-\mu_s\bigr)
\bigl(a_i[\ell]-\mu_i\bigr)\,
P_{k\ell}\,\Delta a_s\,\Delta a_i.
\label{eq:cov}
\end{align}
Equivalently, one can form the signal marginal \(p_s[k]=\sum_\ell P_{k\ell}\,\Delta a_i\) and compute \(V_s=\sum_k (a_s[k]-\mu_s)^2 p_s[k]\,\Delta a_s\); the two forms are identical on a rectangular grid.

\paragraph{Linear inference (Reid) variance.}
The optimal linear estimator of \(a_i\) from \(a_s\) in the mean-square sense is
\begin{equation}
\hat a_i \;=\; \mu_i + G\,(a_s-\mu_s),
\qquad
G \;=\; \frac{C_{si}}{V_s}.
\label{eq:linear_estimator}
\end{equation}
Its mean-square error defines the (linear) inferred variance,
\begin{equation}
\mathrm{Var}(a_i\,|\,a_s)_{\mathrm{lin}}
\;=\; V_i \;-\; \frac{C_{si}^2}{V_s}.
\label{eq:reid_inferred}
\end{equation}
In our implementation, Eqs.~\eqref{eq:means}–\eqref{eq:reid_inferred} are evaluated directly from the normalized table \(P_{k\ell}\) generated from either \(I_q\) (far field) or \(I_x\) (near field). The corresponding \emph{inferred standard deviations} are \(\Delta a_{i|s}=\sqrt{\mathrm{Var}(a_i\,|\,a_s)_{\mathrm{lin}}}\). For EPR/steering tests, we report products Reid product \cite{reid1989demonstration} such as \( U_x = \Delta x_{i|s}\,\Delta q_{x,i|s}\) and \(U_y = \Delta y_{i|s}\,\Delta q_{y,i|s}\), with the Heisenberg benchmark set by the Fourier convention. 
\begin{figure*}[htbp]
  \centering
  \includegraphics[width=0.9\textwidth]{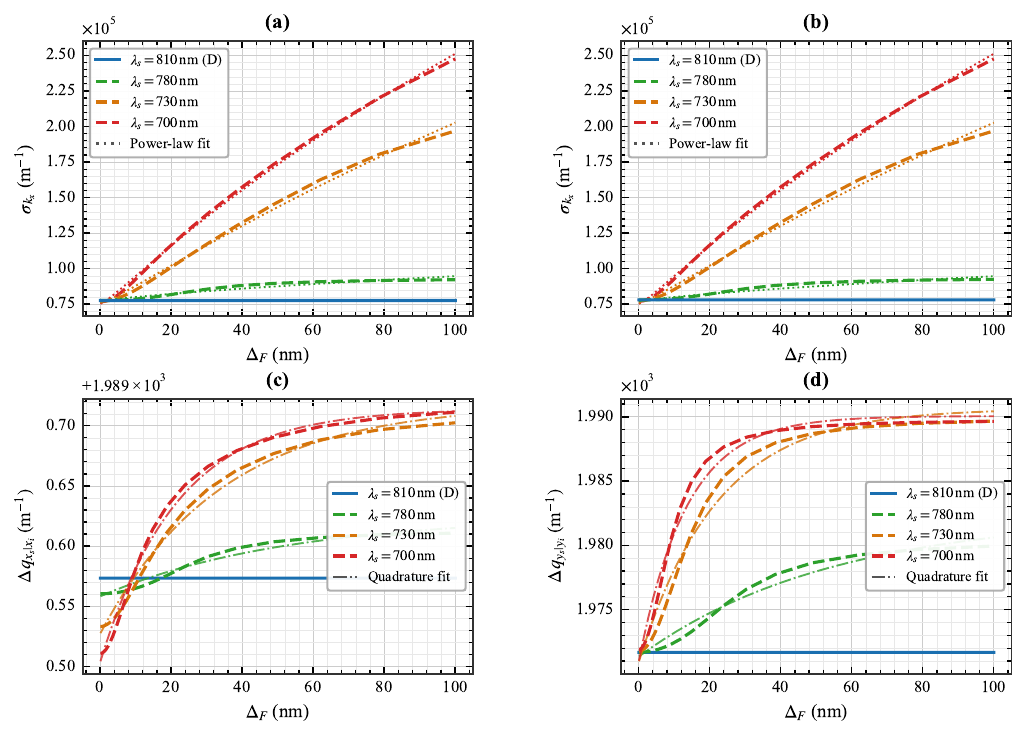}
  \caption{Far-field momentum-space correlations as functions of spectral filter bandwidth for Type-I BBO ($\lambda_p = 405$\,nm, $w_0 = 500\,\mu$m unless varied). Left column: non-walk-off ($x$) axis; right column: walk-off ($y$) axis. (a),(b) Marginal momentum width $\sigma_{k_x}$ and $\sigma_{k_y}$ versus filter FWHM ($\Delta_F$) for the degenerate configuration ($\lambda_s = 810$\,nm, blue solid) and three non-degenerate signal wavelengths (780, 730, 700\,nm). Non-degenerate configurations exhibit monotonic broadening following the power-law crossover model such as Eq.~\ref{eq:marginal_model} (dotted lines, $\beta \approx 1.2$--$1.35$), with the most non-degenerate configuration ($\lambda_s = 700$\,nm) reaching more than twice its narrowband value at $\Delta_F = 100$\,nm. (c), (d) Conditional momentum width $\Delta q_{x,s|i}$ and $\Delta q_{y,s|i}$ versus $\Delta_F$ for $L = 3$\,mm. On both axes the degenerate width is pump-limited and filter-invariant at $\Delta q \approx 1/w_0 = 2000$\,m$^{-1}$. Non-degenerate configurations show a monotonic increase with filter bandwidth, following an exponential saturation model Eq.~\ref{eq:cond_sat} (dashed lines). The walk-off axis (d) shows substantially larger absolute growth and a lower baseline than the non-walk-off axis (c), reflecting the anomalous narrowing of the conditional momentum width by the birefringent walk-off spatial filter at narrow bandwidths and its progressive relaxation as the filter widens.}
  \label{Fig: fig2}
\end{figure*}
\section{Results}
\subsection{Parameter Variations Along the \texorpdfstring{$x$}{x} and 
\texorpdfstring{$y$}{y} Directions in SPDC}

Non-degenerate SPDC is fundamentally distinguished from the degenerate case by a non-zero spatio-spectral coupling $d\theta/d\lambda$. At degeneracy ($\lambda_s = \lambda_i = 810$\,nm), the phase matching angle is at its maximum and $d\theta/d\lambda \approx 0$ by symmetry,
so all wavelengths within a finite filter bandwidth are phase-matched to essentially the same emission angle. In non-degenerate operation, this symmetry is broken and $d\theta/d\lambda \neq 0$, meaning that
different spectral components within the filter passband contribute near-field patterns at different emission angles.

This contrast is directly visible in the X-entanglement pattern of Fig.~\ref{fig:fig1}. Reading off the angular sweep $\Delta\theta$
over a 1\,nm signal bandwidth, the degenerate case
($\lambda_s = \lambda_i = 810$\,nm) gives
$\Delta\theta \approx 0.0002^\circ$, while the non-degenerate example ($\lambda_s = 780$\,nm, $\lambda_i = 842$\,nm) gives $\Delta\theta \approx 0.0358^\circ$ — about $\times 180$ times higher.  This 180-fold difference in spatio-spectral coupling strength underlies all of the qualitative distinctions between degenerate and non-degenerate spatial correlations discussed in this work: the broadening of the marginal momentum distribution, the flat-dip-rise profile in the near field.

In birefringent media such as BBO, the $x$- and $y$-directions (where $y$ denotes the walk-off axis throughout this work) display noticeably different marginal ($\sigma_k$) and conditional momentum ($\Delta q_{s|i}$) distributions. As shown in Fig.~\ref{Fig: fig2} (a, b), our simulations reveal that the marginal distribution (denoted $V$ in Eq.~\ref{eq:Cond_var}) broadens with increasing spectral filter bandwidth ($\Delta_F$, FWHM of filter bandwidth), and that this dependence is present solely in the non-degenerate case. The underlying cause is the strong spatio-spectral coupling that arises in non-degenerate SPDC through the sinc phase matching function of the nonlinear crystal. We evaluate the spatially dependent SPDC parameter--the emission angle $\theta$ and its rate of change with wavelength $d\theta/d\lambda$-- to understand this phenomena. The marginal momentum width $\sigma_{k}$ is well described by the empirical scaling law

\begin{equation}
\sigma_k(\Delta\lambda) =
\sigma_0\sqrt{1 + \left(\frac{\Delta_F}{\Delta\lambda_c}\right)^\beta},
\label{eq:marginal_model}
\end{equation}
where $\Delta\lambda_c = \sigma_0/\alpha \times 2\sqrt{2\ln 2}$, $\alpha = (d\theta/d\lambda)\, k$ and $\beta \approx 1.2–1.35$. We expect $\beta=2$  for Gaussian approximation. $\sigma_0$ in the equation is intrinsic marginal width often estimated using $\sigma_0 = \sqrt{k/L}$, $k = k_s$ for signal arm. We obtained $\sigma_0 = 76,647\, m^{-1}$ from the simulation compared to $\sigma_0 = 70,561 m^{-1}$ with analytical formulae, a error of $7$\%. This difference is because the intrinsic marginal width $\sigma_0$ is approximated by $\sigma_0 \approx \sqrt{k_s/L}$ in the Gaussian phase matching approximation, which underestimates the simulated value by 8--16\% depending on wavelength due to the sinc-shaped tails of the actual phase matching function. Eq.~\eqref{eq:marginal_model} fits the
simulation data to  $<3$\% RMS for all wavelengths studied. The degenerate case ($\lambda_s = 810$\,nm) is flat to within numerical precision ($< 0.003$\%) across the entire filter bandwidth range, confirming that $\frac{\mathrm{d}\theta}{\mathrm{d}\lambda} \approx 0$ at degeneracy completely decouples spectral filtering from the marginal spatial distribution. The crossover FWHM separating the flat and growing regimes is observed at the threshold bandwidth $\Delta\lambda_c$.

Unlike the marginal width, which reflects the angular spread of the accepted emission, the conditional width is governed by transverse momentum conservation: $q_s + q_i \approx q_\mathrm{p}$, which is enforced by the pump beam waist radius as $1/w_0$, independently of which emission angles the spectral filter selects. This distinction leads to qualitatively different behaviour of the marginal and conditional widths under spectral filtering.

We find that at the narrowest filter bandwidths, all four wavelength configurations give $\Delta q \approx 1989$--$1990$\,m$^{-1}$ on both axes, consistent with the theoretical prediction $\Delta q_{\mathrm{s|i}} \approx 1/w_0 = 2000$\,m$^{-1}$ to overall within 0.5\%. The degenerate case (810\,nm) gives $\Delta q =
1989.57$\,m$^{-1}$, which remains strictly constant across the entire filter range $0.1$--$100$\,nm, confirming again that $d\theta/d\lambda \approx 0$ at degeneracy completely decouples the conditional momentum correlation from spectral filtering. The non-degenerate cases show a lower baseline on the walk-off ($y$) axis compared to the non-walk-off ($x$) axis, consistent with the anomalous narrowing reported by Fedorov et al. \cite{fedorov2007anisotropy}: the walk-off term $-(q_s+q_i)\tan\rho$ in the biphoton phase-mismatch imposes an additional constraint on the sum momentum $q_s + q_i$, tightening the conditional distribution beyond the pump-waist limit.

Although the conditional momentum width is essentially pump-limited, a residual growth with filter bandwidth is observed for non-degenerate configurations. The growth follows an exponential saturation law:

\begin{equation}
\Delta q(\Delta_F) =
q_0 + A\!\left(1 - e^{-\Delta_F/\tau}\right),
\label{eq:cond_sat}
\end{equation}
where $q_0 \approx 1/w_0$ is the pump-limited baseline, $A$ is the saturation amplitude representing the maximum possible growth, and $\tau$ is the characteristic bandwidth at which 63\% of the total growth has been reached. This model fits the simulation data with RMS residuals below $0.006$\,m$^{-1}$ for all wavelengths
and both axes. The saturation form is physically motivated: once the filter bandwidth exceeds the spectral range over which the relevant phase matching geometry changes appreciably, no further modification of the conditional distribution occurs and the width approaches a finite asymptote.

The fitted parameters differ markedly between the two transverse axes. On the non-walk-off ($x$) axis as shown in Fig.~\ref{Fig: fig2} (c), the saturation amplitude $A_x$ is at most $0.21$\,m$^{-1}$, representing a fractional growth of less than $0.011$\% of the baseline value. The characteristic bandwidth $\tau_x$ ranges from $22$ to $52$\,nm depending on wavelength, meaning the crossover lies well beyond practical filter bandwidths. For all purposes, the $x$-axis conditional width is filter-invariant, and the small residual growth is attributable to a negligible spectral shift of the centre of the joint momentum distribution rather than
any broadening of the correlation itself.

On the walk-off ($y$) axis, the behaviour is qualitatively different. The saturation amplitude $A_y \approx 10$-$18$\,m$^{-1}$, giving a fractional growth of $0.5$--$0.9$\%, approximately $100$ times larger than the $x$-axis. The characteristic bandwidth $\tau_y$ is correspondingly smaller, with the crossover falling within the $0$--$100$\,nm measurement range visible in Fig.~\ref{Fig: fig2} (d). This enhanced sensitivity arises because the walk-off angle $\rho(\lambda_s)$ is itself wavelength-dependent: as the filter admits spectral slices with different $\rho$, it averages over slightly different walk-off spatial filters on the sum-momentum distribution, progressively relaxing the anomalous narrowing that distinguishes the ND walk-off axis at narrow bandwidths. At large filter bandwidths all non-degenerate curves in Fig.~\ref{Fig: fig2} (d) converge toward the degenerate baseline, confirming this interpretation.

These results demonstrate that spectral filtering modifies the marginal momentum distribution strongly but leaves the conditional momentum correlation with small change for all practical filter bandwidths. These findings can be interpreted as follows: the marginal width reflects which emission angles are selected by the filter, while the conditional width reflects how tightly transverse momentum is conserved between the photon pair, a property governed by the pump beam waist alone. Spectral filtering can arbitrarily alter the former without materially changing the latter, confirming that the EPR momentum correlation in non-degenerate SPDC is robust against broadband detection to within $1$\% even on the more sensitive walk-off axis. The practical implication is that filter bandwidth is not a critical parameter for preserving momentum entanglement — only for preserving position entanglement.

\begin{figure*}[htbp]
  \centering
  \includegraphics[width=0.9\textwidth]{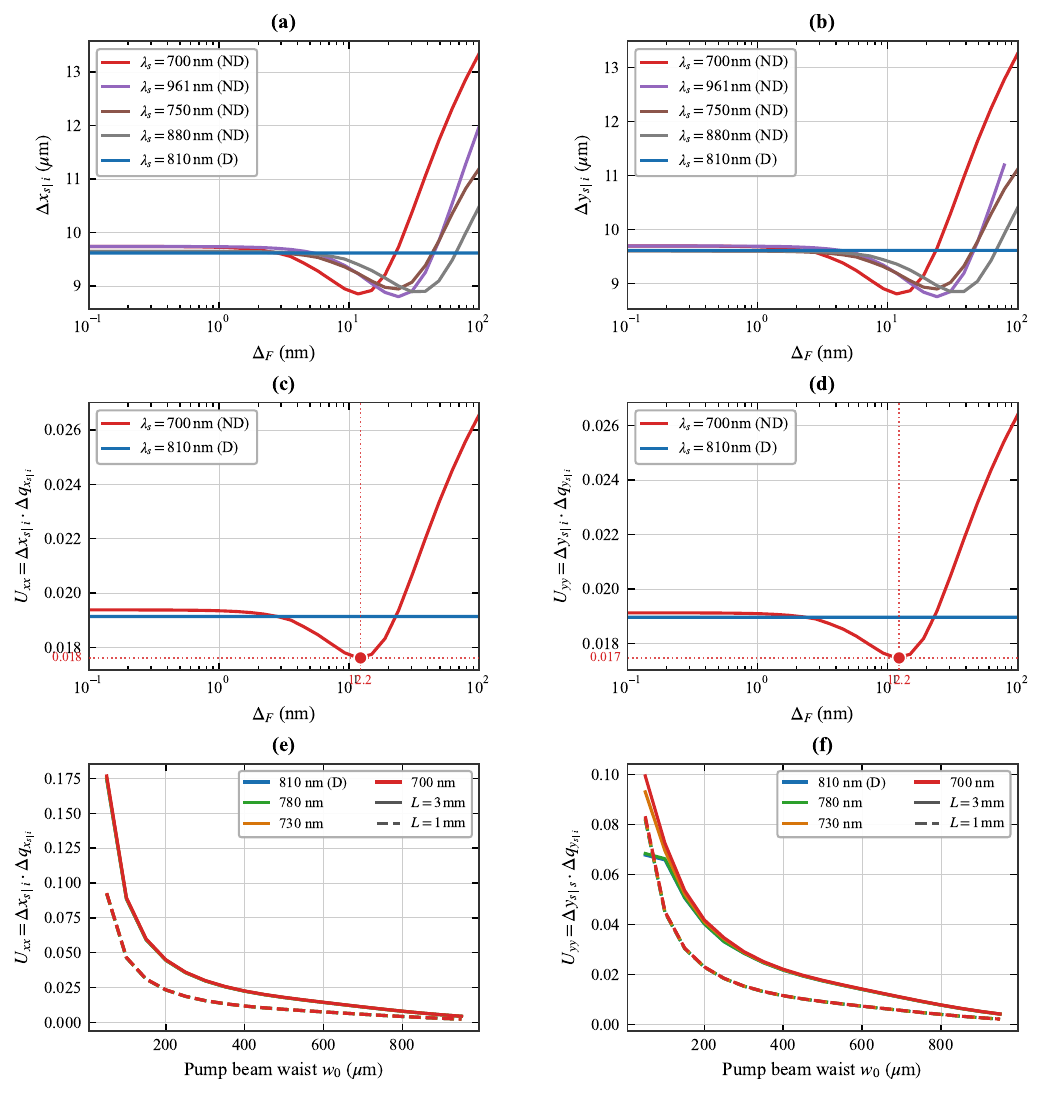}
  \caption{Near-field spatial correlations $\Delta x_{s|i}/\Delta y_{s|i}$ and EPR uncertainty products as functions of spectral filter bandwidth and pump beam waist for Type-I BBO ($\lambda_p = 405$\,nm, $L = 3$\,mm, $w_0 = 500\,\mu$m unless varied). (a),(b) Conditional position width $\Delta x_{s|i}$ and $\Delta y_{s|i}$ versus filter FWHM for the degenerate configuration ($\lambda_s = 810$\,nm, blue) and four non-degenerate signal wavelengths (700\,nm, 750\,nm, 880\,nm) together with the corresponding idler at 961\,nm. All curves exhibit the flat-dip-rise profile: an invariant plateau at narrow bandwidths where $\Delta x_{s|i}^{(0)} \approx 9.5\,\mu$m for all configurations, a minimum near the intrinsic SPDC phase matching bandwidth $\Delta_\text{dip} \approx 1.35\,\Delta\lambda_\text{SPDC}$, and a subsequent rise driven by geometric displacement of spectrally offset near-field slices. The degenerate case remains flat across the entire filter range, consistent with $d\theta/d\lambda = 0$ at degeneracy. The 961\,nm idler curve (violet) exhibits its dip at $(\lambda_i/\lambda_s)^2 \approx 1.88$ times the 700\,nm signal dip location, confirming the exact scaling law of Eq.~(\ref{eq:dip_scaling}). (c),(d) Reid EPR uncertainty product $\Delta x_{s|i}\cdot\Delta q_{x,s|i}$ and $\Delta y_{s|i}\cdot\Delta q_{y,s|i}$ versus filter FWHM for the 700\,nm non-degenerate (red) and 810\,nm degenerate (blue) configurations. Values below O.5 certify spatial entanglement. The red filled circle marks the optimal filter bandwidth $\Delta_F = \Delta_\text{dip}$ at which the uncertainty product reaches its minimum, yielding approximately 10\% improvement over the single-slice limit. The non-degenerate uncertainty product rises steeply beyond $\Delta_\text{dip}$, while the degenerate product remains filter-invariant. (e),(f) Reid uncertainty products $\Delta x_{s|i}\cdot\Delta q_{x,s|i}$ and $\Delta y_{s|i}\cdot\Delta q_{y,s|i}$ versus pump beam waist $w_0$ for $L = 3$\,mm (solid) and $L = 1$\,mm (dashed) at three non-degenerate signal wavelengths and the degenerate case. In all configurations the uncertainty product decreases monotonically with increasing $w_0$, confirming the pump-waist-limited regime $\Delta q_{s|i} \approx 1/w_0$. The walk-off ($y$) axis consistently yields a lower uncertainty product than the non-walk-off ($x$) axis due to the anomalous narrowing of the conditional momentum width by the birefringent walk-off spatial filter, a feature absent in periodically poled quasi-phase-matched crystals.}
  \label{fig:entanglement_certification}
\end{figure*}

\subsection{Near-field conditional position width}
\label{sec:nearfield}

The effect of spectral filtering on the conditional position width $\Delta x_{s|i}$ is fundamentally distinct from its effect on the marginal momentum distribution. Rather than monotonic broadening, the near-field conditional width exhibits a three-stage profile as the filter bandwidth $\Delta_F$ increases: an invariant plateau, a narrowing dip, and a subsequent rise--flat-dip-rise profile as shown in Fig.~\ref{fig:entanglement_certification} (a,b).

For narrow filter bandwidths ($\Delta_F \lesssim \Delta\lambda_\text{SPDC}$), all spectral components admitted by the filter are phase-matched to essentially the same emission angle and wavenumber. The conditional position distribution is therefore identical across all components, and the spectrally averaged width is the standard intrinsic conditional width \cite{edgar2012imaging}:
\begin{equation}
    \Delta x_{s|i}^{(0)} \approx \sqrt{\frac{0.488L\lambda_p}{2\pi}},
\end{equation}
which gives $\Delta x_{s|i}^{(0)} \approx 9.7\,\mu\text{m}$ for $L = 3$\,mm, consistent with our simulation results. Crucially, this baseline value is approximately equal for the degenerate and all non-degenerate configurations tested, confirming that the near-field conditional width is governed by the crystal length and pump wavelength alone, independent of the signal--idler wavelength splitting (the standard $\sqrt{L}$ scaling of $\Delta x_{s|i}$ is described in appendix \ref{app_D}). For the degenerate case ($\lambda_s = \lambda_i= 810$\,nm), $d\theta/d\lambda = 0$ and the profile remains at this plateau for all filter bandwidths.

\begin{figure}
    \centering
    \includegraphics[width=1\linewidth]{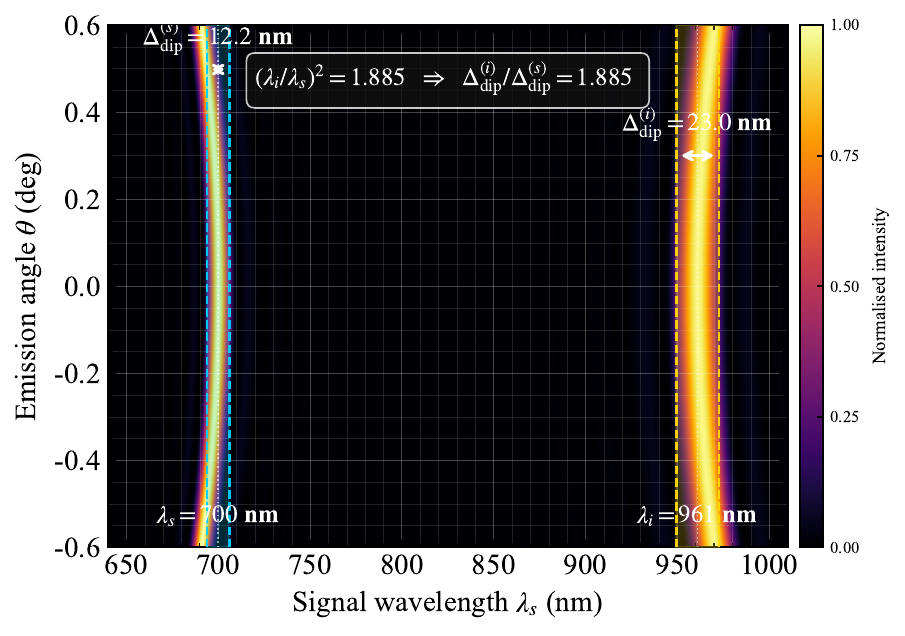}
    \caption{Normalised phase matching intensity X-entanglement (emission angle $\theta$ versus signal wavelength $\lambda_s$), computed for the non-degenerate pump orientation showing signal stripe near 700\,nm and the idler stripe near 961\,nm. dashed vertical lines mark the dip bandwidth at the signal arm, $\Delta_\text{dip}^{(s)} = 12.2$\,nm, centred at 700\,nm; gold dashed lines mark the corresponding idler dip bandwidth, $\Delta_\text{dip}^{(i)} = 23.0$\,nm, centred at 961\,nm. Shaded regions and double-headed arrows indicate the spectral extent of each strip. The ratio $\Delta_\text{dip}^{(i)}/\Delta_\text{dip}^{(s)} = (\lambda_i/\lambda_s)^2 = 1.885$ (inset, bottom right) confirms the exact energy-conservation scaling law to within 0.2\%.}
    \label{fig:X_shape}
\end{figure}

As the filter bandwidth increases beyond $\Delta\lambda_\text{SPDC}$, the conditional width decreases below the intrinsic width. This counterintuitive narrowing arises from a Fourier-width mechanism intrinsic to the non-degenerate phase matching geometry. Shorter-wavelength spectral components within the filter passband have a
larger signal wavenumber $k_s(\lambda_s)$, which broadens the phase matching sinc function in difference-momentum space, $\Delta q_{s|i}$. Via the Fourier relationship $\Delta x_{s|i} \propto 1/\Delta q_{s|i}$, these components produce a narrower near-field conditional distribution. As the filter widens to include these shorter-wavelength components with appreciable weight, they pull the spectrally averaged conditional width below $\Delta x_{s|i}^{(0)}$, producing the dip. We found, the minimum is reached at the dip bandwidth:
\begin{equation}
    \Delta_\text{dip} \approx 1.35\,\Delta\lambda_\text{SPDC},
    \label{eq:dip_bw}
\end{equation}
where $\Delta\lambda_\text{SPDC}$ is the intrinsic SPDC phase matching bandwidth at the signal wavelength, given by
\begin{equation}
    \Delta\lambda_{\text{SPDC}} =
    \frac{0.886\,\dfrac{2\pi}{L}}
    {\left|\dfrac{\mathrm{d}k_s}{\mathrm{d}\lambda_s}
    + \dfrac{\mathrm{d}k_i}{\mathrm{d}\lambda_i}
    \left(\dfrac{\lambda_i}{\lambda_s}\right)^2\right|},
    \label{eq:delta_spdc}
\end{equation}
with the denominator representing the group velocity mismatch i.e., $\mathrm{GVM} = \left|\dfrac{\mathrm{d}k_s}{\mathrm{d}\lambda_s} + \dfrac{\mathrm{d}k_i}{\mathrm{d}\lambda_i} \left(\dfrac{\lambda_i}{\lambda_s}\right)^2\right|$, between signal and idler. The factor of 1.35 arises from the $sinc^2$ SPDC phase matching function. Physically, $\Delta_{dip}$ is the spectral bandwidth over which the central SPDC emission angle has not yet displaced appreciably; it appears in the phase matching X-entanglement spectral width \cite{gatti2009xentanglement} of the bright sinc$^2$ stripe at the signal wavelength, as marked in Fig.~\ref{fig:X_shape} for $\lambda_s = 700$\,nm. The dip minimum corresponds to an improvement of approximately 10\% in the conditional position width relative to intrinsic width, which translates directly into an equivalent improvement in spatial resolution for quantum imaging applications.
For $\Delta_F > \Delta_\text{dip}$, the conditional position width rises above the intrinsic value. This rise has a different physical origin from the dip: it is driven by the geometric displacement of the near-field joint intensity peak. Because $d\theta/d\lambda \neq 0$ for non-degenerate SPDC, spectral slices at longer wavelengths have their near-field peaks displaced from the measurement origin by $\delta x \approx (d\theta/d\lambda)\,
k_s\,(L/2)\,\delta\lambda$. When the filter is wide enough to include slices whose displacement exceeds $\Delta x_{s|i}^{(0)}$, they broaden the spectrally averaged conditional distribution beyond the single-slice limit. The rise is steeper for extreme
non-degenerate configurations (e.g.\ $\lambda_s = 700$\,nm) and becomes progressively shallower closer to degeneracy ($\lambda_s = 780$\,nm), directly reflecting the hierarchy $|d\theta/d\lambda|_{700} \gg |d\theta/d\lambda|_{780}$. This three-stage behaviour is observed identically on both the $x$ and
$y$ transverse axes, confirming that the flat-dip-rise profile is a fundamental property of the non-degenerate phase matching geometry rather than an artefact of the birefringent walk-off.

When the bandpass filter is placed on the idler arm rather than the signal arm, the dip shifts to a larger bandwidth by the exact factor
\begin{equation}
    \Delta_\text{dip}^{(i)} = \left(\frac{\lambda_i}{\lambda_s}\right)^2
    \Delta_\text{dip}^{(s)},
    \label{eq:dip_scaling}
\end{equation}
confirmed numerically to within 0.2\% for all configurations tested. This follows from energy conservation, which requires $d\lambda_i/d\lambda_s = -(\lambda_i/\lambda_s)^2$: the group velocity mismatch per unit bandwidth is smaller at the longer idler wavelength by precisely this factor, so a proportionally wider idler filter is needed to reach the same dip condition-- as also observed in Fig.~\ref{fig:X_shape}, the higher-frequency (shorter-wavelength) signal exhibits a narrower X-entanglement stripe than the lower-frequency (longer-wavelength) idler. Equivalently, the same scaling holds for the spatio-spectral coupling slopes, $(\lambda_i/\lambda_s)^2 = |d\theta/d\lambda|_s / |d\theta/d\lambda|_i$, (see appendix \ref{sec:spatio_spec_coup} for more details)
confirming a consistent underlying mechanism. In practice, this means that a wider bandpass filter is required on the longer-wavelength (lower-energy) idler arm to achieve the same reduction in conditional position uncertainty — a consideration of direct practical importance for non-degenerate quantum imaging systems where the signal and idler are detected in separate arms. 

The optimal filter bandwidth for minimum conditional position uncertainty is $\Delta_F = \Delta_\text{dip}$, given by Eq.~(\ref{eq:dip_bw}) and (\ref{eq:delta_spdc}). Filters narrower than $\Delta_\text{dip}$ sacrifice the $\approx 10\%$ narrowing benefit; filters wider than $\Delta_\text{dip}$ progressively degrade the conditional position correlation due to the geometric
displacement mechanism described above.
\subsection{Entanglement certification via EPR uncertainty products}
Spatial entanglement in photon pairs is commonly certified through the EPR criterion, which tests whether the product of conditional position and momentum uncertainties falls below the bound imposed by the Heisenberg uncertainty principle \cite{reid1989demonstration, Schneeloch2016TransverseSPDC,howell2004realization}. Specifically, a bipartite state is EPR-entangled if
\begin{equation}
    U = \Delta x_{s|i}\cdot\Delta q_{x,s|i} < \tfrac{1}{2}.
    \label{eq:reid}
\end{equation}
This condition does not violate the Heisenberg uncertainty principle, because the position and momentum measurements are performed on different particles \cite{Schneeloch2016TransverseSPDC}. Rather, it constitutes a rigorous certification of spatial entanglement: the tighter the product, the stronger the correlations.

The three-stage flat-dip-rise profile of $\Delta x_{s|i}$ in near-field regime discussed in the previous Sec.~\ref{sec:nearfield} is directly reflected in the EPR uncertainty product. Since the conditional momentum width $\Delta q_{x,s|i}$ remains essentially pump-limited and filter-invariant across all practical bandwidths (Sec.~\ref{sec:momentum}), the variation in $U$ is driven almost entirely by the near-field conditional position width. Accordingly, $U$ attains its minimum at the optimal filter bandwidth $\Delta_F = \Delta_\text{dip}$, as shown in Figs.~\ref{fig:entanglement_certification} (c) and
\ref{fig:entanglement_certification}(d) for x and y directions, respectively presented in cases of $\lambda_s = 700$\,nm (non-degenerate) and $\lambda_s = 810$\,nm (degenerate). For the non-degenerate case, selecting $\Delta_F = \Delta_\text{dip}$ yields an approximately 10\% reduction in $U$ relative to the
single-wavelength limit, a direct consequence of the Fourier-width narrowing mechanism identified in Sec.~\ref{sec:nearfield}. Beyond $\Delta_\text{dip}$, $U$ rises steeply as geometric displacement of spectrally offset near-field slices degrades the position correlation. The degenerate case ($\lambda_s = 810$\,nm) produces a flat, filter-invariant $U$ throughout, again due to the condition $d\theta/d\lambda \approx 0$ condition.

The behaviour is qualitatively identical on both the $x$ and $y$ transverse axes. However, the walk-off axis consistently yields a slightly lower absolute value of $U_{yy} < U_{xx}$. This asymmetry arises from the anomalous narrowing of the conditional momentum width on the walk-off axis — the birefringent walk-off imposes an additional spatial filter on the sum-momentum distribution, tightening
$\Delta q_{y,s|i}$ below the pump-waist-limited value
$1/w_0$ \cite{fedorov2007anisotropy}. This walk-off induced filtering benefit is absent in periodically poled quasi-phase-matched SPDC, where the crystal symmetry eliminates the walk-off term, and represents a distinctive advantage of bulk birefringent crystals for applications requiring strong walk-off-axis correlations.

Figure~\ref{fig:entanglement_certification}(e,f) shows the uncertainty products $U_{xx}$ and $U_{yy}$ as functions of pump beam waist $w_0$ for crystal lengths $L = 1$\,mm and $L = 3$\,mm, at three non-degenerate signal wavelengths and the degenerate configuration. The $U$ decreases monotonically with increasing $w_0$, scaling as $U \propto 1/w_0$, consistent with the pump-waist-limited conditional momentum width $\Delta q_{s|i} \approx 1/w_0$ and independent of $L$. This scaling is well established in the literature \cite{Schneeloch2016TransverseSPDC,fedorov2007anisotropy} and is confirmed here across both degenerate and non-degenerate configurations. A shorter crystal ($L = 1$\,mm, dashed) produces a lower $U$ than a
longer one ($L = 3$\,mm, solid) at the same pump waist, because the conditional position width scales as $\Delta x_{s|i} \propto \sqrt{L}$ (see in Sec.~\ref{app_C} in appendix) while the conditional momentum width is $w_0$-limited and $L$-independent. This consistent with experimentally established results reported in Ref.~\cite{brambila2025certifying} and establishes that crystal length and pump waist are independently
tunable parameters for optimising entanglement strength (the details of $\Delta x_{s|i} \propto \sqrt{L}$ dependence produced using our simulation is given in Sec.~\ref{app_D} in appendix). All configurations tested in this work satisfy $U < 0.5$ across the full waist range, confirming robust spatial entanglement certification in bulk Type-I BBO for both degenerate and non-degenerate SPDC and walk-off and non-walk-off axes. The walk-off axis product $U_{yy}$ falls
below $U_{xx}$ throughout, owing to the spatial filtering effect of birefringent walk-off \cite{fedorov2007anisotropy}, reinforcing the conclusion that bulk birefringent crystals offer a structural enhancement of entanglement on the walk-off axis that is inaccessible in quasi-phase-matched devices.

Taken together, these findings demonstrate that spatial entanglement in bulk non-degenerate SPDC is a tunable quantity, controllable through the choice of filter bandwidth, pump beam waist, and crystal length. The optimal strategy for maximising entanglement certification is to select $\Delta_F = \Delta_\text{dip}$, use the largest practicable pump waist, and choose the shortest crystal consistent
with sufficient pair generation rate for the intended application.

\section{Discussion}
\label{sec:discussions}
The results presented in this work establish a comprehensive picture of how spectral filtering, pump beam waist, and crystal length jointly govern spatial entanglement in bulk Type-I BBO across degenerate and
non-degenerate SPDC configurations.

Our simulation confirms three complementary results in the large-pump regime ($w_0 \gg \sqrt{L/k_\text{eff}}$ on the non-walk-off axis and
$w_0 \gg \rho L/2$ on the walk-off axis). First, the conditional position widths are approximately equal on both axes, $\Delta x_{s|i} \approx \Delta y_{s|i}$, consistent with Da Costa Moura and Monken \cite{dacosta2024epr} and Patil et al.\ \cite{patil2023anisotropic}, and both follow the same $\sqrt{L}$ scaling governed by $k_\text{eff}$. Second, the conditional momentum width is smaller on the walk-off axis, $\Delta q_{y,s|i} < \Delta
q_{x,s|i}$, because the walk-off term $(q_s+q_i)\tan\rho$ imposes an additional constraint on the sum momentum beyond the pump-envelope limit \cite{fedorov2007anisotropy}. Third, combining these results, $U_{yy} < U_{xx}$ in the large-pump regime. This ordering reverses in the focused-pump regime $w_0 \sim l_t = \rho L/2$ \cite{dacosta2024epr}, and the walk-off enhancement is absent entirely in quasi-phase-matched geometries where crystal symmetry eliminates the walk-off term. These findings confirm the established $\sqrt{L}$ scaling of the conditional position width \cite{Schneeloch2016TransverseSPDC, edgar2012imaging, brambila2025certifying} on both axes, and the $1/w_0$ scaling of the conditional momentum width, independently of crystal length.

The central result of this work is the flat-dip-rise profile of the near-field conditional position width as a function of spectral filter bandwidth, observed for all non-degenerate configurations on both transverse axes. This behaviour has no counterpart in prior studies of bulk BBO or ppKTP.

Although demonstrated here for Type-I BBO, the underlying mechanism is general. The three conditions required for the profile to appear are: (i) a finite crystal length $L$ defining $\Delta\lambda_\text{SPDC}$; (ii) non-zero spatio-spectral coupling $d\theta/d\lambda \neq 0$; and
(iii) incoherent spectral averaging. None of these is specific to Type-I BBO — they hold for Type-II phase matching, ppKTP, ppLN, waveguide, and fibre-based SPDC equally.

The dip location $\Delta_\text{dip} \approx 1.35\,\Delta\lambda_\text{SPDC}$ scales as $1/L$: a longer crystal narrows $\Delta\lambda_\text{SPDC}$
and pushes the dip to smaller filter bandwidths, making the effect most easily observed in short crystals at strongly non-degenerate wavelength pairs where GVM is large. The dip depth of $\approx 10\%$
depends on the crystal dispersion profile; larger GVM yields a sharper dip, while quasi-phase-matched sources gain an additional degree of freedom to engineer the dip location via the poling period.

The scaling law $\Delta_\text{dip}^{(i)}/\Delta_\text{dip}^{(s)} =
(\lambda_i/\lambda_s)^2$ follows from energy conservation alone and is independent of crystal material, phase-matching type, or wavelength. It provides a parameter-free experimental test performable on any non-degenerate SPDC source using two narrowband filters and a camera. The degenerate exception is equally universal: at degeneracy,
GVM $= 0$ by symmetry, so $\Delta\lambda_\text{SPDC} \rightarrow \infty$ and the flat-dip-rise profile is entirely absent. This categorical distinction between degenerate and non-degenerate operation is independent of the crystal or geometry used, and
experimentalists operating beyond $\Delta_\text{dip}$ will degrade the near-field position correlation and therefore the achievable imaging resolution.
\section{Conclusion}
\label{sec:conclusion}
We have presented a systematic investigation of spatial entanglement certification in bulk Type-I BBO for both degenerate and non-degenerate SPDC, with full treatment of the effects of pump beam waist, crystal length, and spectral filter bandwidth on transverse spatial correlations. The main findings are as follows.

In the momentum space, the marginal width grows monotonically with filter bandwidth for non-degenerate configurations, following a power-law crossover with exponent $\beta \approx 1.2$--$1.35$
determined by the sinc-shaped phase matching profile. The conditional momentum width is pump-limited and filter-invariant to within 0.01\% on the non-walk-off axis, confirming that far-field entanglement correlations are robust against broadband detection.

In the near-field position space, spectral filtering produces a previously unreported flat-dip-rise profile for all non-degenerate wavelength configurations. The dip minimum occurs at the filter
bandwidth $\Delta_\text{dip} \approx 1.35\,\Delta\lambda_\text{SPDC}$, delivering approximately 10\% improvement in the conditional position width relative to the narrowband limit. When the filter is placed on the idler arm, the dip shifts to $(\lambda_i/\lambda_s)^2$ times the signal-arm location, a parameter-free scaling law confirmed to within 0.2\% and providing a direct experimental signature of spatio-spectral coupling via energy conservation. For the degenerate configuration,
both the marginal broadening and the near-field dip are absent, since $d\theta/d\lambda = 0$ at degeneracy fully decouples spectral filtering from spatial correlations.

The Reid EPR uncertainty product is smaller on the walk-off axis than on the non-walk-off axis throughout the large-pump regime, a structural feature of bulk birefringent crystals arising from
walk-off-induced narrowing of the conditional momentum width. This asymmetry is absent in quasi-phase-matched geometries and represents a distinguishing advantage of bulk BBO for walk-off-axis entanglement. In all configurations studied, $U < 0.5$ is satisfied across the full range of pump waists and crystal lengths, with $U \propto 1/w_0$
confirmed in both degenerate and non-degenerate cases.

Taken together, these results provide the first analytical filter selection guidelines for non-degenerate quantum imaging in bulk
birefringent crystals: the optimal strategy is to operate at $\Delta_F = \Delta_\text{dip}$, use the largest practicable pump beam waist, and select the shortest crystal consistent with the required
photon pair generation rate. The framework presented here generalises straightforwardly to other bulk Type-I phase-matched crystals and provides a quantitative basis for designing filter configurations in ghost imaging, quantum illumination, and entanglement-enhanced sensing systems. 

\appendix

\section{Phase matching}
\label{app_A}

Spontaneous parametric down-conversion (SPDC) is a nonlinear optical process in which a pump photon with frequency $\omega_p$ and wavenumber
$\mathbf{k}_p$ is converted into two lower-frequency photon pairs — the signal and idler — with frequencies $\omega_s$, $\omega_i$ and wavenumbers $\mathbf{k}_s$, $\mathbf{k}_i$, written as
\begin{align}
  \omega_p &= \omega_s + \omega_i,
  \label{eq:energy_conservation} \\
  \mathbf{k}_p &\approx \mathbf{k}_s + \mathbf{k}_i,
  \label{eq:momentum_conservation}
\end{align}
where phase matching admits both perfect and near-perfect phase matching. The residual
phase-mismatch is estimated using
\begin{equation}
    \Delta\mathbf{k} = \mathbf{k}_p - \mathbf{k}_s - \mathbf{k}_i,
\end{equation}
which must be satisfied simultaneously in the longitudinal and transverse directions:
\begin{align}
  \Delta k_z &= k_{pz} - k_{sz} - k_{iz}, \nonumber \\
  \Delta k_x &= k_{px} - q_{sx} - q_{ix}, \\
  \Delta k_y &= k_{py} - q_{sy} - q_{iy}. \nonumber
  \label{eq:deltaks}
\end{align}
The component expansions follow from the SPDC emission geometry:
$k_{pz} = k_p\cos\rho$, $k_{sz} = k_s\cos\theta_s$,
$k_{iz} = k_i\cos\theta_i$, and
$q_{sx} = k_s\sin\theta_s\cos\phi_s$,
$q_{ix} = k_i\sin\theta_i\cos\phi_i$,
$q_{sy} = k_s\sin\theta_s\sin\phi_s$,
$q_{iy} = k_i\sin\theta_i\sin\phi_i$,
where $\rho$ is the walk-off angle of the extraordinarily polarised pump, $\theta_{s(i)}$ are the polar emission angles, and $\phi_{s(i)}$
are the azimuthal emission angles with respect to the pump axis. The longitudinal wavenumber components are
\begin{equation}
    k_{sz(iz)} = \sqrt{k_{s(i)}^2 - q_{sx(ix)}^2 - q_{sy(iy)}^2}.
\end{equation}
In the paraxial approximation ($\theta \ll 1$), the longitudinal phase-mismatch reduces to
\begin{equation}
\Delta k_z = k_{z_p} - k_{z_s} - k_{z_i}
             - \frac{q_s^2 + q_i^2}{2k_p}
             - (q_s+q_i) \tan\rho,
\label{eq:DeltaKz}
\end{equation}
where $q_x^2 = q_{x_s}^2 + q_{x_i}^2$ and
$q_y^2 = q_{y_s}^2 + q_{y_i}^2$.

For Type-I SPDC in BBO the pump is extraordinarily polarised and the signal and idler are ordinarily polarised, so the pump refractive index depends on the angle $\theta_p$ between the pump propagation direction and the crystal optic axis:
\begin{equation}
    \frac{1}{n_e(\theta_p,\lambda)^2}
    = \frac{\cos^2\theta_p}{n_e^2(\lambda)}
    + \frac{\sin^2\theta_p}{n_o^2(\lambda)}.
    \label{eq:n_eff}
\end{equation}
The phase matching angle is found by solving
\begin{equation}
\theta_p
= \cos^{-1}\!\left[
\left(
    \frac{\lambda_s^2\lambda_i^2}
         {\lambda_p^2(\lambda_s n_i + \lambda_i n_s)^2}
    - \frac{1}{(n_p^e)^2}
\right)
\left(
    \frac{(n_p^o n_p^e)^2}
         {(n_p^e)^2 - (n_p^o)^2}
\right)
\right]^{1/2},
\label{eq:phasematching_angle}
\end{equation}
and the walk-off angle for Type-I SPDC is
\begin{equation}
    \rho = \arctan\!\left[
    \frac{\sin(2\theta_\text{pm})}{2}
    \cdot
    \frac{\left(\dfrac{n_o^2}{n_e^2} - 1\right)}
         {1 + \left(\dfrac{n_o^2}{n_e^2} - 1\right)\cos^2\theta_\text{pm}}
    \right].
    \label{eq:walk-off}
\end{equation}
The phase matching efficiency is
\begin{equation}
    \eta \propto \operatorname{sinc}^2\!\left(\frac{\Delta k_z L}{2}\right),
    \label{eq:eta}
\end{equation}
where $L$ is the crystal length. The efficiency peaks at $\Delta k = 0$ and the bandwidth in $k$-space scales as $2\pi/L$, so a longer crystal yields narrower phase matching bandwidth and stronger spatial correlations.

\subsection*{Sellmeier equations for BBO}

The ordinary and extraordinary refractive indices of BBO as a function of wavelength $\lambda$ (in \textmu m) are given by the Sellmeier
equations \cite{dmitriev2013handbook}:
\begin{align}
n_o^2(\lambda) &= 2.7359
    + \frac{0.01878}{\lambda^2 - 0.01822}
    - 0.01354\,\lambda^2,
\label{eq:sellmeier_no} \\[4pt]
n_e^2(\lambda) &= 2.3753
    + \frac{0.01224}{\lambda^2 - 0.01667}
    - 0.01516\,\lambda^2.
\label{eq:sellmeier_ne}
\end{align}
These expressions are valid over the transparency range $0.19$--$3.5\,\mu$m. For $\lambda_p = 405$\,nm the phase matching angle from Eq.~(\ref{eq:phasematching_angle}) is
$\theta_p = 28.82^\circ$ for the degenerate case
($\lambda_s = \lambda_i = 810$\,nm), in agreement with tabulated values.

\subsection*{Spatio-spectral coupling: derivation of
\texorpdfstring{$d\theta/d\lambda_s$}{dtheta/dlambda}}
\label{sec:spatio_spec_coup}

A key quantity governing how spectral filtering affects spatial correlations in non-degenerate SPDC is the rate at which the phase-matched emission angle $\theta$ changes with signal wavelength,
$|d\theta/d\lambda_s|$. We refer to this as the spatio-spectral coupling strength.

Starting from the collinear phase matching condition
$k_p(\theta_p) = k_s(\lambda_s) + k_i(\lambda_i)$ with $\lambda_i = \lambda_s\lambda_p/(\lambda_s - \lambda_p)$ from energy conservation, the crystal orientation $\theta_p$ is fixed by the experimental setup. As $\lambda_s$ varies around the phase-matched
value, the emission angle $\theta$ of the signal photon must adjust to maintain the sinc$^2$ phase matching condition. Implicit differentiation of the phase matching condition with respect to $\lambda_s$ gives
\begin{equation}
    \frac{d\theta}{d\lambda_s}
    = \frac{1}{k_s\cos\theta}
      \left(\frac{dk_s}{d\lambda_s}
            + \frac{dk_i}{d\lambda_i}\frac{d\lambda_i}{d\lambda_s}\right),
    \label{eq:dtheta_dlam_general}
\end{equation}
where the idler wavelength derivative follows from energy conservation:
\begin{equation}
    \frac{d\lambda_i}{d\lambda_s}
    = -\left(\frac{\lambda_i}{\lambda_s}\right)^2.
    \label{eq:dlambda_i}
\end{equation}
The wavenumber dispersions are obtained from the Sellmeier equations (\ref{eq:sellmeier_no})--(\ref{eq:sellmeier_ne}):
\begin{equation}
    \frac{dk_{s(i)}}{d\lambda_{s(i)}}
    = -\frac{2\pi}{\lambda_{s(i)}^2}
      \left(n_o(\lambda_{s(i)})
            + \lambda_{s(i)}\frac{dn_o}{d\lambda}\bigg|_{\lambda_{s(i)}}
      \right).
    \label{eq:dkdlam}
\end{equation}
Combining Eqs.~(\ref{eq:dtheta_dlam_general})--(\ref{eq:dkdlam}), the denominator of Eq.~(\ref{eq:dtheta_dlam_general}) is also the group velocity mismatch (GVM) between signal and idler:
\begin{equation}
    \mathrm{GVM}
    \equiv \left|\frac{dk_s}{d\lambda_s}
           + \frac{dk_i}{d\lambda_i}
             \left(\frac{\lambda_i}{\lambda_s}\right)^2\right|,
    \label{eq:GVM}
\end{equation}
which enters directly into the intrinsic SPDC phase matching bandwidth:
\begin{equation}
    \Delta\lambda_\text{SPDC}
    = \frac{0.886\,(2\pi/L)}{\mathrm{GVM}}.
    \label{eq:delta_spdc_app}
\end{equation}
The intrinsic SPDC phase matching bandwidth, defined as the spectral range of signal wavelengths over which the sinc$^2$ phase matching function remains within its first half-power points at a fixed emission angle, is given by
\begin{equation}
    \Delta\lambda_{\text{SPDC}}
    = \frac{0.886\,(2\pi/L)}{\mathrm{GVM}},
    \qquad
    \mathrm{GVM} = \left|
    \frac{dk_s}{d\lambda_s}
    + \frac{dk_i}{d\lambda_i}
      \left(\frac{\lambda_i}{\lambda_s}\right)^2
    \right|,
    \label{eq:delta_spdc}
\end{equation}
where the factor $0.886 = 2 \times 0.443$ follows from the sinc$^2(x) = 0.5$ condition at $x = 0.443$.
This expression is valid for non-degenerate configurations ($\lambda_s \neq \lambda_i$). At degeneracy, $\lambda_s = \lambda_i$ implies $\mathrm{GVM} = 0$ and therefore $\Delta\lambda_\text{SPDC} \rightarrow \infty$: all signal wavelengths in a neighbourhood of the degenerate frequency satisfy the phase matching condition simultaneously, so no finite spectral filter can resolve any phase matching structure. This is the fundamental reason why the conditional position width is filter-invariant in the degenerate case — there is no spatio-spectral coupling to exploit or degrade.
\begin{figure*}[htbp]
  \centering
  \includegraphics[width=0.9\textwidth]{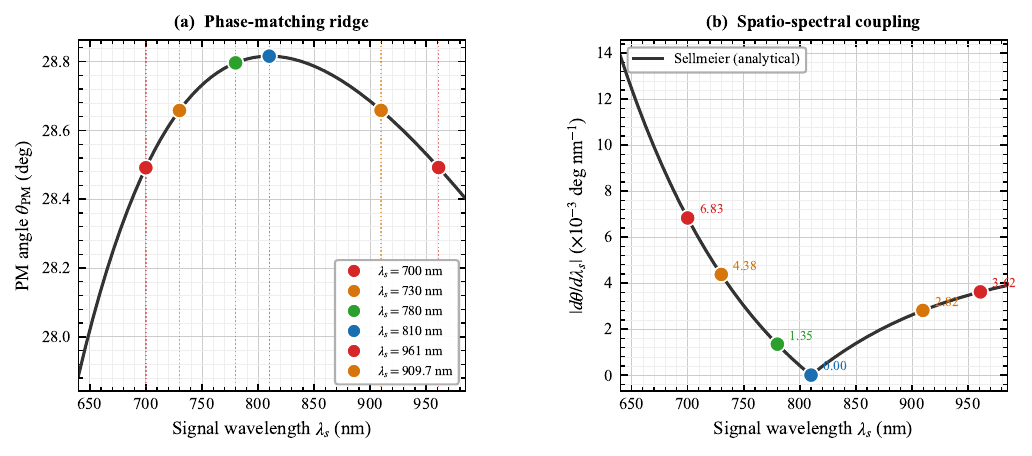}
  \caption{Phase matching geometry and spatio-spectral coupling strength for Type-I BBO ($\lambda_p = 405$\,nm, $L = 3$\,mm).
(a) phase matching angle $\theta_\text{PM}$ versus signal wavelength $\lambda_s$, computed analytically from the BBO Sellmeier equations Eq.~\ref{eq:phasematching_angle}. (b) Spatio-spectral coupling strength $|d\theta/d\lambda_s|$ versus
signal wavelength, derived analytically from the Sellmeier equations. The coupling vanishes at the degenerate wavelength ($\lambda_s = 810$\,nm) where $d\theta/d\lambda = 0$ by symmetry.}
  \label{fig:dtheta_dlambda}
\end{figure*}
Fig.~\ref{fig:dtheta_dlambda} (a) shows the phase-matching angle for four wavelength configurations, and Fig.~\ref{fig:dtheta_dlambda} (b) shows its gradient $|d\theta/d\lambda_s|$ computed from Eqs.~(\ref{eq:dtheta_dlam_general})--(\ref{eq:dkdlam}) using the BBO Sellmeier equations, plotted as a function of signal wavelength. The coupling strength vanishes at degeneracy ($\lambda_s = 810$\,nm)
and grows approximately linearly with the non-degeneracy parameter $\varepsilon = |\lambda_s - \lambda_i|/(\lambda_s + \lambda_i)$,
following $|d\theta/d\lambda_s| \propto \varepsilon^{1.05}$. The values at the four degeneracy conditions studied are summarised in Table~\ref{tab:dtheta}.

\begin{table}[h]
\centering
\caption{Intrinsic SPDC bandwidth
$\Delta\lambda_\text{SPDC}$, and predicted dip bandwidth $\Delta_\text{dip} \approx 1.35\,\Delta\lambda_\text{SPDC}$ for
$L = 3$\,mm and $\lambda_p = 405$\,nm.}
\label{tab:dtheta}
\begin{tabular}{ccccccc}
\hline\hline
$\lambda_s$ & $\lambda_i$ & $\varepsilon$ & $\Delta\lambda_\text{SPDC}$ &
$\Delta_\text{dip,s}$ & $\Delta_\text{dip,i}$ &
$\Delta_\text{dip,i}/\Delta_\text{dip,s}$ \\
(nm) & (nm) & & (nm) & (nm) & (nm) &  \\
\hline
700 & 961 & 0.157 & 9.0 & 12.2 &  23 & 1.88 \\
730 & 910 & 0.110 & 14.1 & 19.0 & 29.5 & 1.55 \\
750 & 880 & 0.110 & 17.7 & 23.9 & 32.9 & 1.357 \\
780 & 842 & 0.038 & 45.7 & 61.7 & 72.13 & 1.17 \\
810 & 810 & 0.000 & $\infty$ & $\infty$ & $\infty$ & undefined \\
\hline\hline
\end{tabular}
\end{table}
\section{Pump beam waist effect}
\label{app_B}

\begin{figure*}[htbp]
  \centering
  \includegraphics[width=1\textwidth]{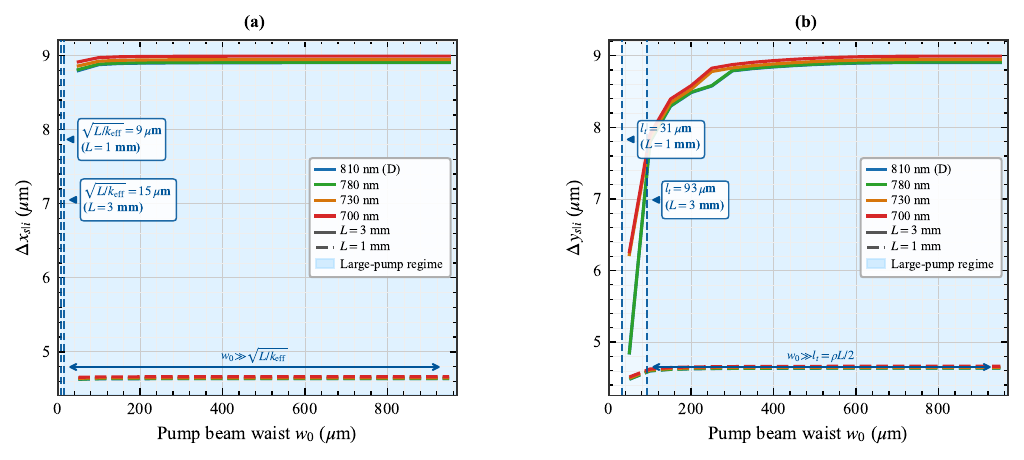}
  \caption{Effect of pump beam waist variation on conditional position correlation for (a) x-direction and (b) y-direction for $L = 1\, mm$ and $L = 3\,mm$ for $\Delta_F = 10\,nm$.}
  \label{fig:positionCor_pumpwaist}
\end{figure*}

We assume the pump beam has a Gaussian spatial profile:
\begin{equation}
    E(q_\perp) = \exp\!\left[-\frac{w_0^2 q_\perp^2}{4}\right],
    \label{eq:pump_waist}
\end{equation}
where $w_0$ is the pump beam waist and
$q_\perp^2 = q_x^2 + q_y^2$ is the squared transverse wavenumber.
The biphoton angular spectrum is then
\begin{equation}
    \Phi(q_x,q_y)
    \propto \exp\!\left[-\frac{(q_x^2+q_y^2)w_0^2}{4}\right]
    \operatorname{sinc}^2\!\left(\frac{\Delta k_z L}{2}\right),
    \label{eq:phasematching_function}
\end{equation}
which encodes the two fundamental length scales governing SPDC spatial correlations: the pump beam waist $w_0$ and the crystal length $L$. Fig.~\ref{fig:positionCor_pumpwaist} shows the pump beam waist size as a function of conditional position uncertainty, where at large pump beam waist $w_0\gg\sqrt{L/k}$, showed $\Delta x_{s|i}$ is invariant to pump waist in the x axis. A small initial increase is observed in the walk-off axis because in the walk-off axis the large pump waist regime is set by  $w_0>>l_t=\rho L/2$, which is $\approx 93\, \mu m$ with $\rho = 3.5^\circ$.

\section{Conditional momentum width and \texorpdfstring{$1/w_0$}{1/w0} scaling}
\label{app_C}

The conditional momentum width $\Delta q_{s|i}$ characterises how tightly transverse momentum conservation $q_s + q_i \approx q_\text{pump}$ is enforced between signal and idler. Since this constraint is imposed by the pump angular spectrum,
$\Delta q_{s|i}$ is set exclusively by the pump beam waist and is independent of crystal length in the paraxial regime:
\begin{equation}
    \Delta q_{s|i} \approx \frac{1}{w_0}.
    \label{eq:cond_mom}
\end{equation}

\begin{figure*}[htbp]
  \centering
  \includegraphics[width=0.9\textwidth]{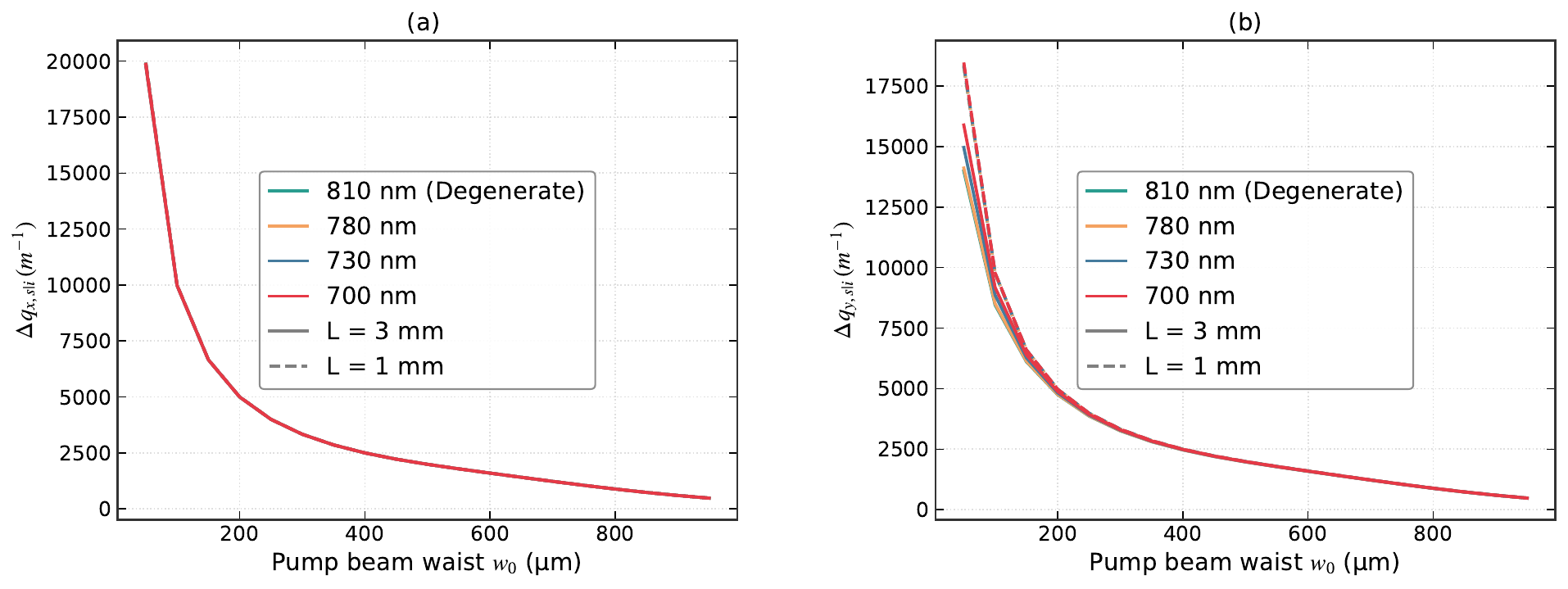}
  \caption{Effect of pump beam waist variation on conditional momentum correlation for (a) x-direction and (b) y-direction for $L = 1\, mm$ and $L = 3\,mm$, and $\Delta_F = 10\,nm$.}
  \label{fig:cond_mom_waist}
\end{figure*}

Fig.~\ref{fig:cond_mom_waist} confirms this scaling on both the non-walk-off ($x$) and walk-off ($y$) axes across all wavelength configurations and crystal lengths studied [$L = 1$\,mm (dashed) and $L = 3$\,mm (solid)].
Two features are evident. First, on the non-walk-off ($x$) axis [Fig.~\ref{fig:cond_mom_waist} (a)], $\Delta q_{x,s|i}$ follows $1/w_0$ precisely for all wavelengths and both crystal lengths, confirming that the conditional momentum correlation on this axis is governed entirely by the pump envelope and is insensitive to crystal length or signal--idler wavelength splitting. All curves collapse onto
a single universal line, consistent with Eq.~(\ref{eq:cond_mom}) and
with the predictions of Schneeloch and Howell
\cite{Schneeloch2016TransverseSPDC} and Howell et al.\ \cite{howell2004realization}.

Second, on the walk-off ($y$) axis
[Fig.~\ref{fig:cond_mom_waist} (b)], the scaling remains $\Delta q_{y,s|i} \propto 1/w_0$ but the absolute values are systematically lower than on the $x$-axis, and the non-degenerate configurations ($\lambda_s = 700$\,nm in particular) show a clear
separation from the degenerate case at small pump waists. This asymmetry arises from the walk-off spatial filter on the sum-momentum distribution: the term $-(q_s+q_i)\tan\rho$ in the biphoton phase-mismatch [Eq.~(\ref{eq:DeltaKz})] constrains
$q_{y,s} + q_{y,i}$ beyond the pump-envelope limit, tightening the conditional momentum distribution on the walk-off axis \cite{fedorov2007anisotropy, dacosta2024epr}. The $L$-dependence of
$\Delta q_{y,s|i}$ — absent on the $x$-axis — reflects the progressive strengthening of this walk-off filter as the interaction length increases.
\section{Conditional position width and \texorpdfstring{$\sqrt{L}$}{sqrt(L)} scaling}
\label{app_D}

In the near-field (position space), the conditional position width $\Delta x_{s|i}$ is governed by the phase matching sinc function rather than the pump envelope. Schneeloch and Howell \cite{Schneeloch2016TransverseSPDC} showed that in the single-wavelength (narrowband) limit, the conditional position width follows:
\begin{equation}
    \Delta x_{s|i}
    \approx \sqrt{\frac{0.488L\lambda_p}{2\pi}}
    = \sqrt{\frac{0.488L}{k_\text{eff}}},
    \label{eq:sqrt_L_scaling}
\end{equation}
where the effective wavenumber for non-degenerate SPDC is
\begin{equation}
    k_\text{eff}
    = \frac{2k_sk_i}{k_s + k_i},
    \label{eq:k_eff}
\end{equation}
the harmonic mean of the signal and idler wavenumbers. In the degenerate limit $k_s = k_i = k$ this reduces to $k_\text{eff} = k = 2\pi n/\lambda_{s/i}$, recovering the standard result. For $\lambda_p = 405$\,nm and $L = 3$\,mm this gives
$\Delta x_{s|i} \approx 9.5\,\mu$m, consistent with our simulation results.
\begin{figure*}[htbp]
  \centering
  \includegraphics[width=0.9\textwidth]{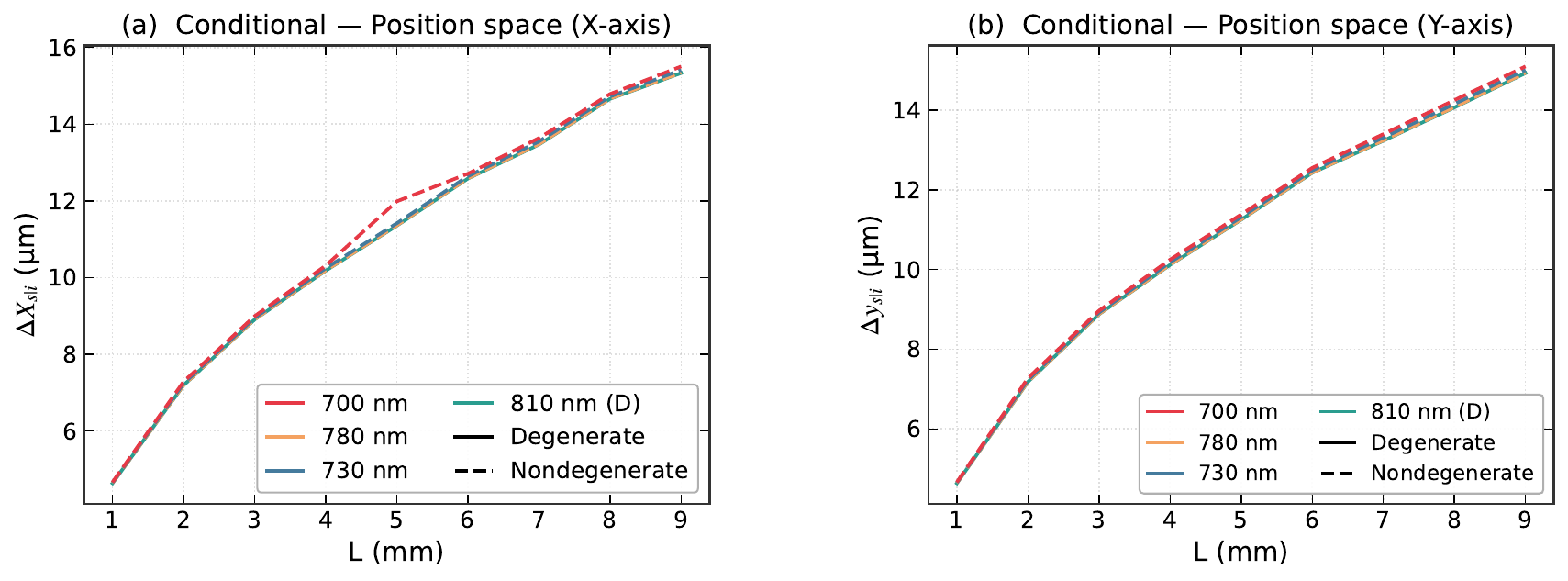}
  \caption{The $\sqrt{L}$ dependence of conditional position correlation in (a) x-axis (b) y-axis.}
  \label{fig:sqrt_L}
\end{figure*}

Fig.~\ref{fig:sqrt_L} shows $\Delta x_{s|i}$ and $\Delta y_{s|i}$ as functions of crystal length $L$ for four wavelength configurations. Several features are evident.

First, both axes follow the same $\sqrt{L}$ scaling, confirming that the conditional position width is governed by the longitudinal phase matching structure independently of birefringent walk-off. The data shows the trend of Eq.~(\ref{eq:sqrt_L_scaling}) with $k_\text{eff}$ computed from the Sellmeier equations, for all wavelengths and on both axes.

Second, the absolute values are nearly identical on the $x$ and $y$ axes for all wavelengths — $\Delta x_{s|i} \approx \Delta y_{s|i}$ — despite the strong momentum-space asymmetry between the axes. This
equality is a consequence of the $\sqrt{L}$ scaling being set by the longitudinal sinc structure, which is isotropic in the paraxial limit, and is consistent with the analytical prediction of Da Costa Moura and
Monken \cite{dacosta2024epr} and the numerical results of Patil et al. \cite{patil2023anisotropic}.

Third, the $\sqrt{L}$ scaling is independent of pump beam waist in the large-pump regime $w_0 \gg \sqrt{L/k_\text{eff}}$, (we have taken pump waist equals $w_0 = 500 \mu m$ for the $\sqrt{L}$ scaling shown here, see Appendix Sec.~\ref{fig:sqrt_L} for further details) confirming that crystal length and pump beam waist independently control the conditional position and momentum widths, respectively. The effect of different pump waist sizes on these parameters is tabulated in Table~\ref{Tab:pump_waist_regime}. This independence is the foundation of the two-parameter design strategy for quantum imaging: $L$ sets the position correlation strength and $w_0$ sets the momentum correlation strength.

\begin{table}[h]
\centering
\small 
\setlength{\tabcolsep}{1.5pt} 
\begin{tabular}{lccc}
\hline
 \textbf{Regime} \\\textbf{(non-walk-off axis)} & \textbf{condition} & \textbf{$\Delta x_{\mathrm{s|i}}$} & \textbf{$\Delta q_{\mathrm{s|i}}$} \\ \hline
Large pump (typical) & $w_0 \gg \sqrt{L/k}$ & $\sqrt{2L/k}$, independent of $w_0$ & $\approx 1/w_0$ \\
Small pump (tight focus) & $w_0 \ll \sqrt{L/k}$ & $\approx w_0$, independent of L & $\approx 1/w_0$ \\
Intermediate & $w_0 \sim \sqrt{L/k}$ & depends on both & $\approx 1/w_0$ \\
\textbf{Regime} \\\textbf{(walk-off axis)} & &  &  \\ \hline
Large pump (typical) & $w_0 \gg \rho L/2$ &  independent of $w_0$ & $\approx 1/w_0$ \\
Small pump (tight focus) & $w_0 \ll \rho L/2$ & $\approx w_0$, independent of L & $\approx 1/w_0$ \\
Intermediate & $w_0 \sim \rho L/2$ & depends on both & $\approx 1/w_0$ \\
\hline
\end{tabular}
\caption{Pump waist regimes for non-walk-off and walk-off axes.}
\label{Tab:pump_waist_regime}
\end{table}
\bibliography{ref} 
\end{document}